\documentclass[runningheads]{llncs}

\usepackage{times}

\usepackage{soul}
\usepackage{url}
\usepackage{graphicx}
\usepackage{amsmath}
\usepackage{booktabs}
\usepackage{outlines}
\usepackage{dirtytalk}
\usepackage{wrapfig}
\usepackage{capt-of}

\usepackage[linesnumbered,ruled,noend]{algorithm2e}
\usepackage{amsfonts}
\newcommand{\norm}[1]{\left\lVert#1\right\rVert}

\renewcommand\vec{\mathbf}

\DeclareMathOperator*{\argmax}{arg\,max}

\begin{document}

\title{EvoBA: An Evolution Strategy as a Strong Baseline for Black-Box Adversarial Attacks}
\titlerunning{EvoBA}

\author{Andrei Ilie \and
Marius Popescu \and
Alin Stefanescu}
\authorrunning{Ilie et al.}
%
\institute{University of Bucharest, Romania \\
\email{\{cilie, marius.popescu, alin\}@fmi.unibuc.ro}}

\maketitle


\begin{abstract}
Recent work has shown how easily white-box adversarial attacks can be applied to state-of-the-art image classifiers. However, real-life scenarios resemble more the black-box adversarial conditions, lacking transparency and usually imposing natural, hard constraints on the query budget.

We propose $\textbf{EvoBA}$ \footnote{All the work is open source: https://github.com/andreiilie1/BBAttacks.}, a black-box adversarial attack based on a surprisingly simple evolutionary search strategy. $\textbf{EvoBA}$ is query-efficient, minimizes $L_0$ adversarial perturbations, and does not require any form of training.

$\textbf{EvoBA}$ shows efficiency and efficacy through results that are in line with much more complex state-of-the-art black-box attacks such as $\textbf{AutoZOOM}$. It is more query-efficient than $\textbf{SimBA}$, a simple and powerful baseline black-box attack, and has a similar level of complexity. Therefore, we propose it both as a new strong baseline for black-box adversarial attacks and as a fast and general tool for gaining empirical insight into how robust image classifiers are with respect to $L_0$ adversarial perturbations.

There exist fast and reliable $L_2$ black-box attacks, such as $\textbf{SimBA}$, and $L_{\infty}$ black-box attacks, such as $\textbf{DeepSearch}$. We propose $\textbf{EvoBA}$ as a query-efficient $L_0$ black-box adversarial attack which, together with the aforementioned methods, can serve as a generic tool to assess the empirical robustness of image classifiers. The main advantages of such methods are that they run fast, are query-efficient, and can easily be integrated in image classifiers development pipelines.

While our attack minimises the $L_0$ adversarial perturbation, we also report $L_2$, and notice that we compare favorably to the state-of-the-art $L_2$ black-box attack, $\textbf{AutoZOOM}$, and of the $L_2$ strong baseline, $\textbf{SimBA}$.

\end{abstract}

\section{Introduction}
With the increasing performance and applicability of machine learning algorithms, and in particular of deep learning, the safety of such methods became more relevant than ever. There has been growing concern over the course of the last few years regarding adversarial attacks, i.e., algorithms which are able to fool machine learning models with minimal input perturbations, as they were shown to be very effective.

Ideally, theoretical robustness bounds should be obtained in the case of critical software involving image classification components. There has been important recent research in this direction \cite{gopinath2018deepsafe,huang2017safety,dvijotham2018dual,ruan2018reachability}, but most often the algorithms generating these bounds work for limited classes of models, do not scale well with larger neural networks, and require complete knowledge of the target model's internals. Therefore, complementary empirical robustness evaluations are required for a better understanding of how robust the image classifiers are. In order to achieve this, one has to come up with effective adversarial attacks that resemble real-life conditions, such as in the black-box  query-limited scenario.

In general, adversarial attacks are classified as either white-box or black-box. White-box adversarial attacks, where the attacker has complete knowledge of the target model, were shown to be particularly successful, most of them using gradient-based methods \cite{szegedy2013intriguing,goodfellow2015explaining,athalye2018obfuscated}. In the case of black-box adversarial attacks, the attacker can only query the model, and has no access to the model internals and to the data used for training. These restrictions make the black-box adversarial setup resemble more real-life scenarios. Furthermore, the attacker usually has to minimise the number of queries to the model, either due to time or monetary constraints (such as in the case of some vision API calls). 

Previous state-of-the-art black-box adversarial attacks focused on exploiting the transferability phenomenon, which allowed the attackers to train substitute models imitating the target one, and perform white-box attacks on these \cite{papernot2016transferability,papernot2016limitations}. More recently, a class of black-box adversarial attacks, called Zeroth Order Optimization ($\textbf{ZOO}$)  \cite{chen2017zoo}, has gained momentum, providing one of the current state-of-the art attacks, $\textbf{AutoZOOM}$ (\cite{tu2019autozoom}). Interestingly, a much simpler algorithm, $\textbf{SimBA}$ (Simple Black-box Attack) \cite{guo2019simple}, achieves a similar, slightly lower success rate than state-of-the-art attacks, including $\textbf{AutoZOOM}$, and requires a lower number of queries. Therefore, $\textbf{SimBA}$ is proposed to be used as a default baseline for any adversarial attack, but it is itself an unexpectedly powerful algorithm.

We propose $\textbf{EvoBA}$, an untargeted black-box adversarial attack that makes use of a simple evolutionary search strategy. $\textbf{EvoBA}$ only requires access to the output probabilities of the target model for a given input and needs no extra training. $\textbf{EvoBA}$ is more query-efficient than $\textbf{SimBA}$ and $\textbf{AutoZOOM}$, and has a perfect success rate, surpassing $\textbf{SimBA}$ and being aligned with $\textbf{AutoZOOM}$ from this point of view. We designed $\textbf{EvoBA}$ to minimize the $L_0$ adversarial perturbation, however we also report the $L_2$ norms of the perturbations it generates and compare them with the $L_2$ norms of $\textbf{AutoZOOM}$ and $\textbf{SimBA}$, which are $L_2$ adversarial attacks. Despite our algorithm aiming to minimize a different metric than these methods, we achieve similar $L_2$ norms in the perturbations we generate, and a significantly better $L_0$.

$\textbf{EvoBA}$ is a model-agnostic empirical tool that can be used as a test to assess how robust image classifier systems are with respect to $L_0$ noise. Therefore, $\textbf{EvoBA}$ is both a simple and strong baseline for black-box adversarial attacks, but also a generic $L_0$ robustness evaluation method that can be used in a fast and reliable way with any system involving image classification.

While $L_2$ (Euclidean distance) and $L_{\infty}$ (maximum absolute pixel-wise distance) adversarial attacks are more commonly studied, $L_0$ (count of modified pixels distance) adversarial attacks can fit better real-life, physical settings, such as in the well-known cases of graffiti perturbations on stop-signs from \cite{eykholt2018robust} and of adversarial eyeglass frames from \cite{sharif2016accessorize}. As \cite{bafna2018thwarting} notes, \say{physical
obstructions in images or malicious splicing of audio or video files are realistic threats that can be
modeled as $L_0$ noise, whereas $L_2$ attacks may be more difficult to carry out in the physical world}.

The main property of adversarially perturbed images is that they are very close from a human point of view to the original benign image, but trick the model into predicting the wrong class. While the notion of \say{closeness from a human perspective} is hard to quantify, there exists general consensus around using the $L_0, L_2, \text{ and } L_{\infty}$ norms as proxies for measuring the adversarial perturbations \cite{wiyatno2019adversarial,papernot2016limitations,papernot2016transferability,guo2019simple,zhang2020deepsearch}. 

$\textbf{EvoBA}$ focuses on the $L_0$ norm, is fast, query-efficient, and effective. Therefore, we propose using it together with similarly fast and efficient methods that focus on different norms, such as $\textbf{SimBA}$, which focuses on $L_2$, and $\textbf{DeepSearch}$ (\cite{zhang2020deepsearch}), which focuses on $L_{\infty}$, to empirically evaluate the robustness of image classifiers. These methods can act together as a fast and general toolbox used along the way of developing systems involving image classification models. $\textbf{EvoBA}$ can easily be incorporated in development pipelines for gaining fast insights into a model's $L_0$ empirical robustness, only requiring access to the classifier's prediction method and to a sample of target images together with their corresponding labels.

Moreover, many adversarial training methods focus on improving the robustness with respect to $L_0, L_2, \text{ and } L_{\infty}$ perturbations \cite{sharif2018suitability}. Therefore, it is important to be able to empirically evaluate how much the robustness of target models improve due to adversarial training with respect to these norms. A toolbox consisting of efficient and reliable attacks such as $\textbf{EvoBA}$, $\textbf{DeepSearch}$, and $\textbf{SimBA}$ can serve this purpose.

$\textbf{EvoBA}$ is most surprising through its dual nature, acting both as a strong and fast black-box adversarial attack, such as $\textbf{SimBA}$, but also achieving results which are in line with much more complex, state-of-the art black-box attacks, such as $\textbf{AutoZOOM}$.

To wrap it up, we propose $\textbf{EvoBA}$ as a standard strong baseline for any black-box adversarial attack and as a tool that can provide empirical insight into how robust image classifiers are. Its main advantages are that it is as effective as state-of-the-art black-box attacks, such as $\textbf{AutoZOOM}$, and more query-efficient. While it is an $L_0$ adversarial attack, it achieves $L_2$ perturbations of similar magnitudes with state-of-the-art $L_2$ black-box attacks and requires no training. Furthermore, $\textbf{EvoBA}$ is highly parallelisable, which allows it to run significantly faster than $\textbf{SimBA}$, the other powerful baseline black-box attack.

$\textbf{Our contributions}$:
\begin{itemize}
    \item We propose $\textbf{EvoBA}$, a fast, query-efficient, and effective black-box $L_0$ adversarial attack, which can be used as a generic test for assessing the $L_0$ robustness of image classifiers.
    \item We show how $\textbf{EvoBA}$ compares favourably to the state-of-the art black-box $L_2$ attack $\textbf{AutoZOOM}$, while being significantly simpler.
    \item We show how $\textbf{EvoBA}$ (an $L_0$ attack) serves a similar purpose to the other strong and simple baselines such as  $\textbf{SimBA}$ (an $L_2$ attack) and $\textbf{DeepSearch}$ (an $L_{\infty}$ attack). These methods are suitable to be run together, as part of a robustness testing toolbox, while developing any image classification model.
\end{itemize}

$\textbf{Paper outline}$:
\begin{itemize}
    \item In Section \ref{sect:related_work} we do a quick literature review around related work and position $\textbf{EvoBA}$ with respect to the presented methods.
    \item In Section \ref{sect:the_method} we introduce $\textbf{EvoBA}$ together with the threat model we use. We provide pseudocode for $\textbf{EvoBA}$ and explain the main ideas behind it. We also analyze its space and time complexities.
    \item In Section \ref{sect:experiments} we present our experimental methodology and setup, provide and analyse the quantitative results of $\textbf{EvoBA}$ in comparison with other methods, and present some qualitative results.
    \item In Section \ref{sect:concl} we wrap-up the main findings and propose future research directions for our work.
\end{itemize}


\section{Related Work}\label{sect:related_work}
While the black-box adversarial attack settings are far more restrictive than the white-box ones, their similarity to real-life scenarios has increasingly brought them into the spotlight of machine learning safety.

One general approach for black-box attacks exploits the transferability property of adversarial examples \cite{papernot2016transferability,papernot2016limitations}. The attacker can train their own substitute model and perform white-box attacks on it, which usually yield good adversarial samples for the target model as well. However, \cite{su2018robustness} showed the limitations of this method, highlighting how not all white-box attacks and not all architectures transfer well.

A class of approaches that do not rely on transferability is based on Zeroth Order Optimization (ZOO), which tries to estimate the gradients of the target neural network in order to generate adversarial examples. One of the early algorithms in this direction, $\textbf{ZOO}$ \cite{chen2017zoo} managed to reach similar success rates with state-of-the-art white-box attacks, producing adversarial samples of comparable visual quality. In general, Zeroth Order Optimization refers to any functional optimization where gradients are not available, so the term has commonly been adapted in the field of black-box adversarial attacks as a general category of methods that estimate the gradients.

The main disadvantage of traditional ZOO-type attacks is that they usually require plenty of queries for approximating the coordinate-wise gradients. $\textbf{AutoZOOM}$ \cite{tu2019autozoom} solved this issue by performing a dimensionality reduction on top of the target image, and then using the ZOO approach in the reduced space. It achieved results that are aligned with previous ZOO state-of-the-art methods, but managed to reduce the query count by up to $93\%$ on datasets such as MNIST, ImageNet, and CIFAR-10. Our method is comparable to $\textbf{AutoZOOM}$ in terms of performance, achieving a similar success rate with a slightly lower query count, but has the main advantage of being considerably simpler. Our approach does not need to estimate gradients at all and, compared to $\textbf{AutoZOOM-AE}$ - the more powerful attack from \cite{tu2019autozoom}, it does not require any form of training or knowledge about the training data. In addition, $\textbf{AutoZOOM}$ demands access to the output probabilities over all classes, unlike our method, which only requires the output class and its probability. While $\textbf{AutoZOOM}$ is an attack that minimizes the $L_2$ perturbations norm, $\textbf{EvoBA}$ is focused on minimising the $L_0$ norm. However we also report the $L_2$ and notice that it is comparable to $\textbf{AutoZOOM}$'s results.

$\textbf{SimBA}$ (Simple Black-box Adversarial Attack, \cite{guo2019simple}) is a very simple strategy that has comparable performance to the significantly more complex $\textbf{AutoZOOM}$. The method randomly iterates through all pixels, perturbing them with fixed noise amounts if this makes the model output a lower probability for the correct class. $\textbf{EvoBA}$ has a similar complexity to $\textbf{SimBA}$, both of them being good candidates for strong adversarial attack baselines. $\textbf{SimBA}$ is mainly powered up by the randomness of choosing which pixel to perturb, and the only information it uses is whether the correct probability decreases. In comparison, $\textbf{EvoBA}$ strikes a better balance between exploration and exploitation, which is common for evolutionary algorithms. This makes $\textbf{EvoBA}$ achieve a slightly better success rate than $\textbf{SimBA}$, with a lower query budget. Similarly to $\textbf{AutoZOOM}$, $\textbf{SimBA}$ is an $L_2$ adversarial attack, while $\textbf{EvoBA}$ focuses on $L_0$. Nevertheless, the average $L_2$ perturbation norm of $\textbf{EvoBA}$ is slightly higher, but comparable to $\textbf{SimBA}$'s results. We also run $\textbf{SimBA}$ and remark that the $L_0$ perturbation norms that $\textbf{EvoBA}$ achieves are significantly better. Furthermore, $\textbf{SimBA}$ does not allow for any kind of parallelisation, while the evolution strategy we use in $\textbf{EvoBA}$ is highly parallelisable and, accordingly, faster.


$\textbf{DeepSearch}$ (\cite{zhang2020deepsearch}) is another simple, yet very efficient black-box adversarial attack, which achieves results in line with much more complex methods. It is an $L_{\infty}$ attack, perturbing with very high probability all the pixels (maximal $L_0$ norm of the perturbations), which makes it incomparable with the $L_0$ attack $\textbf{EvoBA}$. Similarly, while $\textbf{EvoBA}$ optimizes the $L_0$ norm of the perturbations, it most often produces high $L_{\infty}$ distortions, with sometimes near-maximal perturbations for the few chosen pixels. While both $\textbf{DeepSearch}$ and $\textbf{EvoBA}$ have non-complex implementations, $\textbf{EvoBA}$ is conceptually simpler. $\textbf{DeepSearch}$ is based on the idea of linear explanations of adversarial examples (\cite{goodfellow2015explaining}), and exploits three main aspects: it devises a mutation strategy to perturb images as fast as possible, it performs a refinement on top of the earliest adversarial example in order to minimise the $L_{\infty}$ norm, and adapts an existing hierarchical-grouping strategy for reducing the number of queries (\cite{moon2019parsimonious}). Furthermore, $\textbf{EvoBA}$ has the advantage of being highly parallelisable, while $\textbf{DeepSearch}$ is inherently sequential. $\textbf{DeepSearch}$, $\textbf{EvoBA}$, and $\textbf{SimBA}$ are complementary methods that serve the similar purpose of efficient and reliable black-box attacks working under the query-limited scenario, each optimising the produced perturbations under a different norm.



As black-box adversarial attacks are ultimately search strategies in obscured environments, it has been natural to also explore the path of evolutionary algorithms. One notable example is $\textbf{GenAttack}$ \cite{alzantot2019genattack}, an approach that follows the classic pattern of genetic algorithms. While it was developed at the same time with $\textbf{AutoZOOM}$, the authors report similar results for the targeted versions of the two methods, without providing any untargeted attack results. We focused on the untargeted scenario, and our results are also in line with $\textbf{AutoZOOM}$.
$\textbf{GenAttack}$ focuses on minimising the $L_{\infty}$ perturbation norm, and, in expectation, the $L_0$ it achieves is equal to the count of all pixels in the image. In comparison, $\textbf{EvoBA}$ is an $L_0$ attack, achieving considerably small perturbations under this norm. In addition, $\textbf{EvoBA}$ is less complex and more suitable for a strong baseline attack.

A related approach to ours, which also makes use of evolution strategies, is \cite{meunier2019yet}, which tries to minimise the $L_{\infty}$ norm of adversarial perturbations.
It proposes different evolution strategies applied on top of a tiling approach inspired by \cite{ilyas2018prior}, where the authors use a Bandits approach.
The attacks they propose focus on minimising the $L_{\infty}$ norm of the perturbations and the authors do not report any other results regarding different norms. The $L_2$ cost of this approach is not clear, one of the main issues being that it can become rather high. The $L_0$ is equal in general to the number of pixels in the entire image, in comparison with $\textbf{EvoBA}$, which is $L_0$-efficient. Qualitatively, the applied adversarial tiles it generates are easily perceivable by a human, yielding grid-like patterns on top of the target image, while the samples produced by $\textbf{EvoBA}$ are imperceptible (Figure \ref{fig:band}, ImageNet) or look like benign noise (Figure \ref{fig:band_cifar10}, CIFAR-10).

\section{The Method} \label{sect:the_method}

\subsection{Notation and threat model}
We work under black-box adversarial settings, with limited query budget and $L_0$ perturbation norm. We consider the untargeted attack scenario, where an adversary wants to cause perturbation that changes the original, correct prediction of the target model for a given image to any other class.

We denote by $\mathbf{F}$ the target classifier. By a slight abuse of notation, we let $\mathbf{F}(\vec{x)}$ be the output distribution probability of model $\mathbf{F}$ on input image $\vec{x}$ and $\mathbf{F}_{k}(\vec{x})$ be the output probability for class $k$. Then, $\mathbf{F}$ can be seen as a function $\mathbf{F}: \mathbb{I} \mapsto \mathbb{R}^K$, where $\mathbb{I}$ is the image space (a subset of $\mathbb{R}^{h \times w \times c}$) and $K$ is the number of classes.

As we are working under black-box conditions, we have no information about the internals of $\mathbf{F}$, but we have query access to it, i.e., we can retrieve $\mathbf{F}(\vec{x})$ for any $\vec{x} \in \mathbb{I}$. In fact, we will see that for our method we just need access to $\argmax_k{\mathbf{F}_k(\vec{x})}$ and to its corresponding probability.

Let us consider an image $\vec{x}$, which is classified correctly by $\mathbf{F}$. The untargeted attack goal is to find a perturbed version $\vec{\Tilde{x}}$ of $\vec{x}$ that would make 
\begin{equation} \label{formulation:untargeted}
    \argmax_k{\mathbf{F}_k(\vec{\Tilde{x}})} \neq \argmax_k{\mathbf{F}_k(\vec{x})},
\end{equation}
constrained by the query and $L_0$ bounds.

\subsection{The algorithm}

It is usual for black-box attacks to deal with a surrogate optimization problem that tries to find a perturbed version $\vec{\Tilde{x}}$ of $\vec{x}$ that minimizes $\mathbf{F}(\vec{\Tilde{x}})$. This is clearly not equivalent to the formulation at (\ref{formulation:untargeted}), but it often yields good adversarial examples and is easier to use in practice.

In  loose terms, this surrogate optimization problem can be formulated as follows for an image $\vec{x}$ with true label $y$:
\begin{equation} \label{formulation:optimization}
    \min_{\boldsymbol{\delta \in \mathbb{R}^{h \times w \times c}}}\mathbf{F}_y(\vec{x} + \boldsymbol{\delta}), \text{w.r.t. queries} \le Q, \norm{\delta}_0 \le \epsilon.
\end{equation}

In order to tackle (\ref{formulation:optimization}), we adopt a simple evolution strategy that yields results in line with state-of-the-art black-box attacks.

Our method (Algorithm \ref{alg::evoba}) works by iteratively creating generations of perturbed images according to the following process: it selects the fittest individual in each generation (with lowest probability to be classified correctly), starting from the unperturbed image, then samples small batches of its pixels and randomly perturbs them, stopping when the fittest individual is either no longer classified correctly or when one of the constraints no longer holds (when either the query count or the distance become too large).

The function $\textsc{SAMPLE\_PIXELS(PARENT, B)}$ does a random, uniform sample over the pixels of $\textsc{PARENT}$, and returns a list of size at most $B$ (the sampling is done with repetition) containing their coordinates. Its purpose is to pick the pixels that will be perturbed. The function $\textsc{SAMPLE\_VALUES}$ generates pixel perturbed values. For our $L_0$ objective, we let it pick uniformly a random value in the pixel values range.

The algorithm follows a simple and general structure of evolution strategies. The mutation we apply on the best individual from each generation is selecting at most $B$ pixels and assigning them random, uniform values. The fitness of an individual is just the cumulative probability of it being misclassified.


\begin{algorithm}[t]
    \DontPrintSemicolon
      \KwData{black-box model $\mathbf{F}$, image $\vec{x}$, correct class $k$, query budget $Q$, $L_0$ threshold $\epsilon$, pixel batch size $B$, generation size $G$}
      $\textsc{parent} \gets \vec{x}$ \; 
      $\textsc{prediction} \gets \mathbf{F}_k(\vec{x})$ \;
      $\textsc{query\_cnt} \gets 1$ \;
      \While{$\norm{\textsc{parent} - \vec{x}}_0 < \epsilon$ and $\textsc{query\_cnt}<Q$}{
        $\textsc{offspring} \gets [\ ]$ \;
        $\textsc{fitnesses} \gets [\ ]$ \;
        $\textsc{pixels} \gets \textsc{sample\_pixels}( \textsc{parent},B)$ \;
        \For{$\textsc{idx} \leftarrow 1 \dots G$}{
            $\textsc{child} \gets \textsc{parent}$ \;
            \For{$\textsc{pixel} \leftarrow \textsc{pixels}$}{
                $\textsc{child}[\textsc{pixel}] \leftarrow \textsc{sample\_values}()$ \;
            }
            $\textsc{offspring} \gets \textsc{offspring} + [\textsc{child}]$ \;
            $\textsc{pred\_child} \gets \mathbf{F}(\textsc{child})$ \;
            $\textsc{query\_cnt} \gets \textsc{query\_cnt} + 1$ \;
            \If{$\argmax \textsc{pred\_child} \neq k$}{\Return\textsc{child}}
            $\textsc{fitnesses} \gets \textsc{fitnesses} + [1 - \textsc{pred\_child}_k(\textsc{child})]$\;
        
        }
        $\textsc{best\_child} \gets \argmax(\textsc{fitnesses})$\;
        $\textsc{parent} \gets \textsc{offspring}[\textsc{best\_child}]$\;
      }
      \Return{$\textsc{Perturbation Failed}$}
      
    \caption{EvoBA}
    \label{alg::evoba}
\end{algorithm}

One important detail that is not mentioned in the pseudocode, but which allowed us to get the best results, is how we deal with multi-channel images. If the target image has the shape $h \times w \times c$ (height, width, channels), then we randomly sample a position in the $h \times w$ grid and add all of its channels to the $\textsc{PIXELS}$ that will be perturbed. We hypothesise that this works well because of a \say{inter-channel transferability} phenomenon, which allows $\textbf{EvoBA}$ to perturb faster the most sensible zones in all the channels. Note that this yields a cost of $c$ for every grid-sample to the $L_0$ perturbation norm, so in the case of ImageNet or CIFAR-10 images it counts as $3$, and in the case of MNIST it counts as $1$.

The query budget and $L_0$ constraints impose a compromise between the total number of generations in an $\textbf{EvoBA}$ run and the size $G$ of each generation. The product of these is approximately equal to the number of queries, so for a fixed budget we have to strike the right balance between them. The bigger $G$ is, the more we favour more exploration instead of exploitation, which should ultimately come at a higher query cost. The smaller $G$ is, the search goes in the opposite direction, and we favour more the exploitation. As each exploitation step corresponds to a new perturbation, this will result in bigger adversarial perturbations. Furthermore, lacking proper exploration can even make the attack unsuccessful in the light of the query count and $L_0$ constraints.

The batch size $B$ allows selecting multiple pixels to perturb at once in the mutation step. It is similar to the learning rate in general machine learning algorithms: the higher it is, the fewer queries (train steps, in the case of machine learning algorithms) we need, at the cost of potentially missing local optimal solutions.

The space complexity of Algorithm \ref{alg::evoba} is $\mathcal{O}(G \times \text{size}(\vec{x}))$. As the size of $\vec{x}$ is in general fixed, or at least bounded for specific tasks, we can argue that the space complexity is $\mathcal{O}(G)$. However, $\textbf{EvoBA}$ can be easily modified to only store two children at a moment when generating new offspring, i.e., the currently generated one and the best one so far, which makes $\textbf{EvoBA}$'s space complexity $\mathcal{O}(1)$.

The dominant component when it comes to time complexity is given by $\mathbf{F}$ queries. In the current form of Algorithm \ref{alg::evoba}, they could be at most $\min(\frac{\epsilon}{B} \times {G}, Q)$. Assuming that the budget $Q$ will generally be higher, the time complexity would roughly be $\mathcal{O}(\frac{\epsilon}{B} \times {G} \times (\text{query cost of } \mathbf{F}))$. This is merely the sequential time complexity given by the unoptimised pseudocode, however the $G$ $\mathbf{F}$-queries can be batched, yielding much faster, parallelised runs.

\section{Experiments}\label{sect:experiments}
\subsection{Experimental Setup}
We used TensorFlow/Keras for all our experiments. All the experiments were performed on a MacBook with 2,6 GHz 6-Core Intel Core i7, without a GPU.

We ran locally $\textbf{SimBA}$, the strong and simple $L_2$ black-box adversarial attack, and compared $\textbf{EvoBA}$ to both our local $\textbf{SimBA}$ results and reported results from the paper introducing it (\cite{guo2019simple}). 

We also monitored and compared against the $\textbf{AutoZOOM}$ results, but for these we have used different target models, as the main focus was on comparing to $\textbf{SimBA}$ (which already achieves results which are in line with state-of-the-art approaches, such as $\textbf{AutoZOOM}$, being more query-efficient), so we adopted their models. 

We don't do a head-to-head comparison with the efficient $L_{\infty}$ black-box baseline $\textbf{DeepSearch}$, as it is a direct consequence of their approach that they get near-maximal $L_0$ perturbations, while $\textbf{EvoBA}$ aims to optimize the $L_0$ norm. Similarly, $\textbf{EvoBA}$ creates high $L_{\infty}$ perturbations, as it modifies very few pixels with random, possibly big quantities. Therefore, $\textbf{DeepSearch}$ and $\textbf{EvoBA}$ are complementary methods that should be used together, but a direct comparison of their results is not suitable, as the $L_0$ and $L_{\infty}$ objectives are partly contradictory.

We ran multiple experiments over four datasets: MNIST \cite{lecun1998mnist}, CIFAR-10 \cite{krizhevsky2009learning}, and ImageNet \cite{imagenet_cvpr09}.


On MNIST we have been running our experiments on a classic target LeNet architecture \cite{lecun2015lenet}, while $\textbf{SimBA}$ does not report any results on this dataset, and $\textbf{AutoZOOM}$ uses a similar architecture to ours, with additional dropout layers (taken from \cite{carlini2017towards}). However, their target models are trained by default with distillation, which is likely to make perturbations harder. For comparing with $\textbf{SimBA}$ on MNIST, we ran the attack ourselves and aggregated various metrics.

We initially used MNIST to validate $\textbf{EvoBA}$ against a completely random black-box adversarial attack, similar to the one introduced in \cite{ilierobustness}. The purely random strategy iterates by repeatedly sampling a bounded number of pixels from the original target image and changing their values according to a random scheme. While the purely random method introduced in \cite{ilierobustness} achieves surprising results for such a simple approach, it does a very shallow form of exploration, restarting the random perturbation process with each miss (i.e., with each perturbed image that is still classified correctly). We will refer to the completely random strategy as $\textbf{CompleteRandom}$ in the experiments below.

For CIFAR-10, we use a ResNet-50 \cite{he2016deep} target model, similarly to $\textbf{SimBA}$, and we compare to both their reported results and to our local run of their attack. 


For ImageNet, we have used a similar target ResNet-50 to the one used in $\textbf{SimBA}$, while $\textbf{AutoZOOM}$ used InceptionV3 \cite{szegedy2016rethinking}.

We will use the following shorthands in the results below: $\textbf{SR}$ (Success Rate), $\textbf{QA}$ (Queries Average), $\textbf{L0}$ (Average of L0 successful perturbations), $\textbf{L2}$ (Average L2 norm of successful perturbations).

We will refer to $\textbf{EvoBA}$ that perturbs at most $B$ pixels at once and that has generation size $G$ as $\textbf{EvoBA}(B,G)$. We will explicitly mention in each section which thresholds were used for the experiments.

For all the local runs of $\textbf{SimBA}$, we have been using $\epsilon=0.2$ (a hyperparameter specific to $\textbf{SimBA}$), which was also used in the paper \cite{guo2019simple}. We only replicated locally the results of the Cartesian Basis version of $\textbf{SimBA}$, which resembles more an $L_0$ adversarial attack, but which is less efficient than the Discrete Cosine Transform (DCT) version from the paper. Therefore, we will use $\textbf{SimBA-LCB}$ to refer to the local run of SimBA on top of the exact same target models as $\textbf{EvoBA}$, with $\epsilon=0.2$. We will use $\textbf{SimBA-CB}$ to refer to the results of the Cartesian Basis paper results, and $\textbf{SimBA-DCT}$ for the DCT paper results.

In the cases where the $\textbf{AutoZOOM}$ paper \cite{tu2019autozoom} provides data, we will only compare to $\textbf{AutoZOOM-BiLIN}$, the version of the attack which requires no additional training and data, and which is closer to our and $\textbf{SimBA}$'s frameworks.

\subsection{Results on MNIST}

We only experiment with $\textbf{EvoBA}(B,G)$ with $B>1$ on MNIST, and focus on $B=1$ for subsequent experiments, for which we impose an $L_0$ perturbation limit of $100$ and a query threshold of $5000$.

Running $\textbf{SimBA}$ ($\textbf{SimBA-LCB}$) on a local machine with the mentioned specifications, requires an average of $\textbf{93s}$ per MNIST sample. In comparison, all the $\textbf{EvoBA}$ experiments on MNIST took between $\textbf{1.94s}$ and $\textbf{4.58s}$ per sample.

We also report the $\textbf{CompleteRandom}$ results, for which we impose an $L_0$ perturbation limit of $100$ and a query threshold of $5000$, similarly to the constraints we use for $\textbf{EvoBA}$.

We randomly sampled $200$ images from the MNIST test set and ran $\textbf{SimBA-LCB}$, $\textbf{EvoBA}$, and $\textbf{CompleteRandom}$ against the same LeNet model \cite{lecun2015lenet}. For reference, we also add the results of $\textbf{AutoZOOM}$, which are performed on a different architecture. Therefore, the results are not directly comparable with them. $L_0$ data is not available for $\textbf{AutoZOOM}$, but it is usual for ZOO methods to perturb most of the pixels, so it is very likely that the associated $L_0$ is very high.

\begin{table}[h]
    \centering
    \begin{tabular}{ c c c c c} 
 \hline
  & SR & QA & L0 & L2 \\ [0.5ex] 
 \hline
 \textbf{EvoBA(1,10)} & 100\% & 301.4 & 29.32 & 3.69 \\ 
 \textbf{EvoBA(1,20)} & 100\% & 549.4 & 26.88 & 3.58 \\
 \textbf{EvoBA(1,40)} & 100\% & 894.2 & 21.92 & 3.33 \\
 \textbf{EvoBA(2,20)} & 100\% & 312.6 & 30,72 & 3.65 \\
 \textbf{EvoBA(2,30)} & 100\% & 265.4 & 38,6 & 3.89 \\
 \textbf{SimBA-LCB} & 56\% & 196.86 & 48.16 & 2.37 \\ 
 \textbf{AutoZOOM-BiLIN} & 100\% & 98.82 & - & 3.3 \\
 \textbf{CompleteRandom} & 60.5\% & 576.1 & 93.98 & 5.59 \\[0.1ex] 
 \hline
    \end{tabular}
    \caption{MNIST results. $\textbf{SimBA-LCB}$ has a very low success rate in the case of a LeNet architecture. If we take a look at the 56\% images perturbed by $\textbf{EvoBA(1,10)}$ for example, the QA becomes 192.8, which is not apparent from the table, but which is more efficient than $\textbf{SimBA-LCB}$.}
    \label{tab:my_label}
\end{table}

We remark that all the $\textbf{EvoBA}$ configurations have a $100\%$ success rate, and $\textbf{SimBA-LCB}$ achieves $56\%$. While $\textbf{SimBA}$ generally achieves near-perfect success rates on other tasks, one could argue that attacking a simple target model such as LeCun on a relatively easy task such as MNIST is much harder than performing attacks for more intricate target models and tasks. This is a natural trade-off between the complexity of a model and its robustness.
\begin{figure}[ht]
    \centering
    \includegraphics[width=0.8\columnwidth]{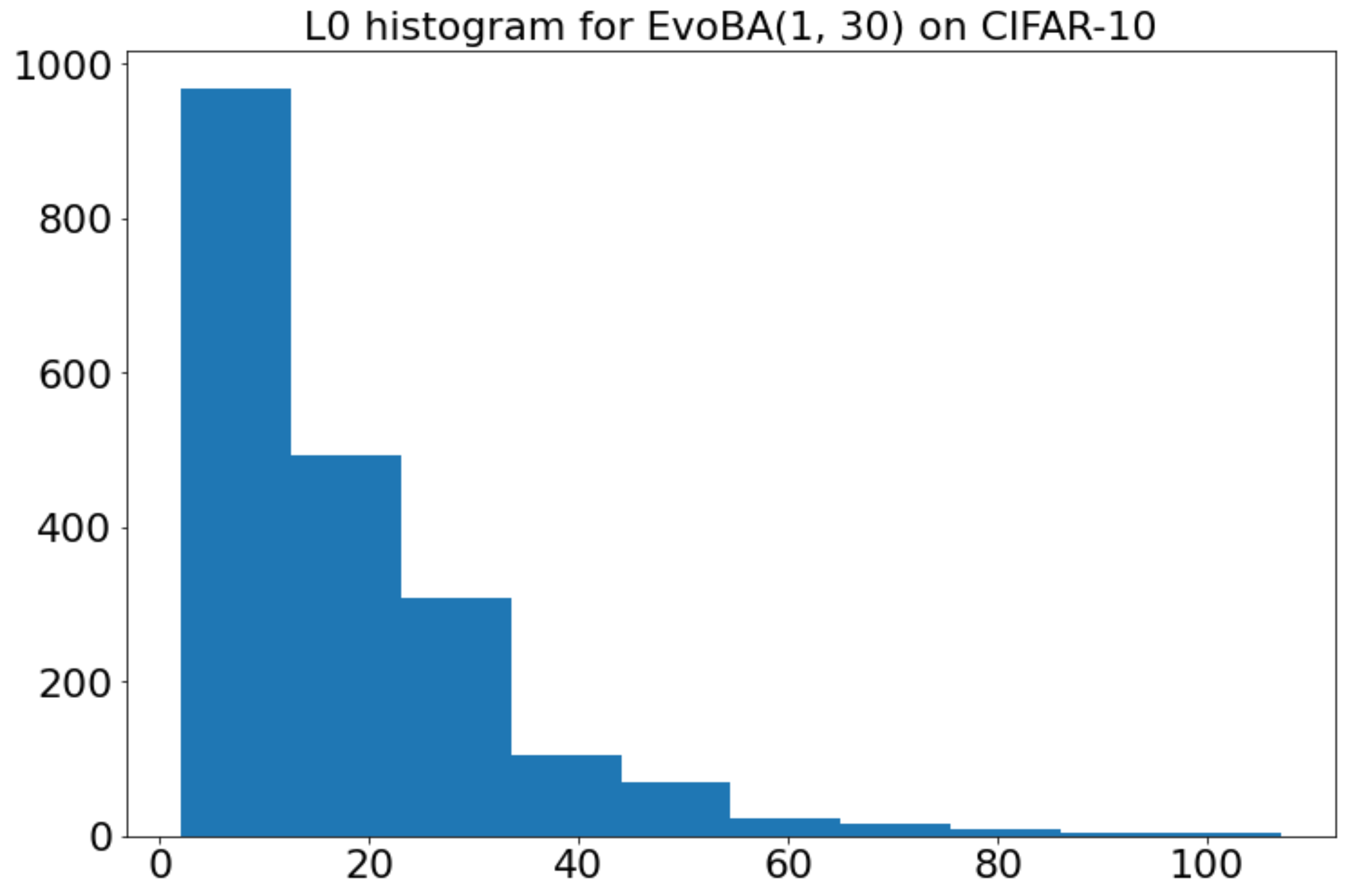}
    \caption{Histogram of $L_0$ perturbation norms obtained by $\textbf{EvoBA(1,30)}$ on CIFAR-10 with target model ResNet-50. The distribution is very heavy on small values, with few outliers.}
    \label{fig:histo_l0_evoba_cifar10}
\end{figure}

\begin{figure}[ht]
    \centering
    \includegraphics[width=0.8\columnwidth]{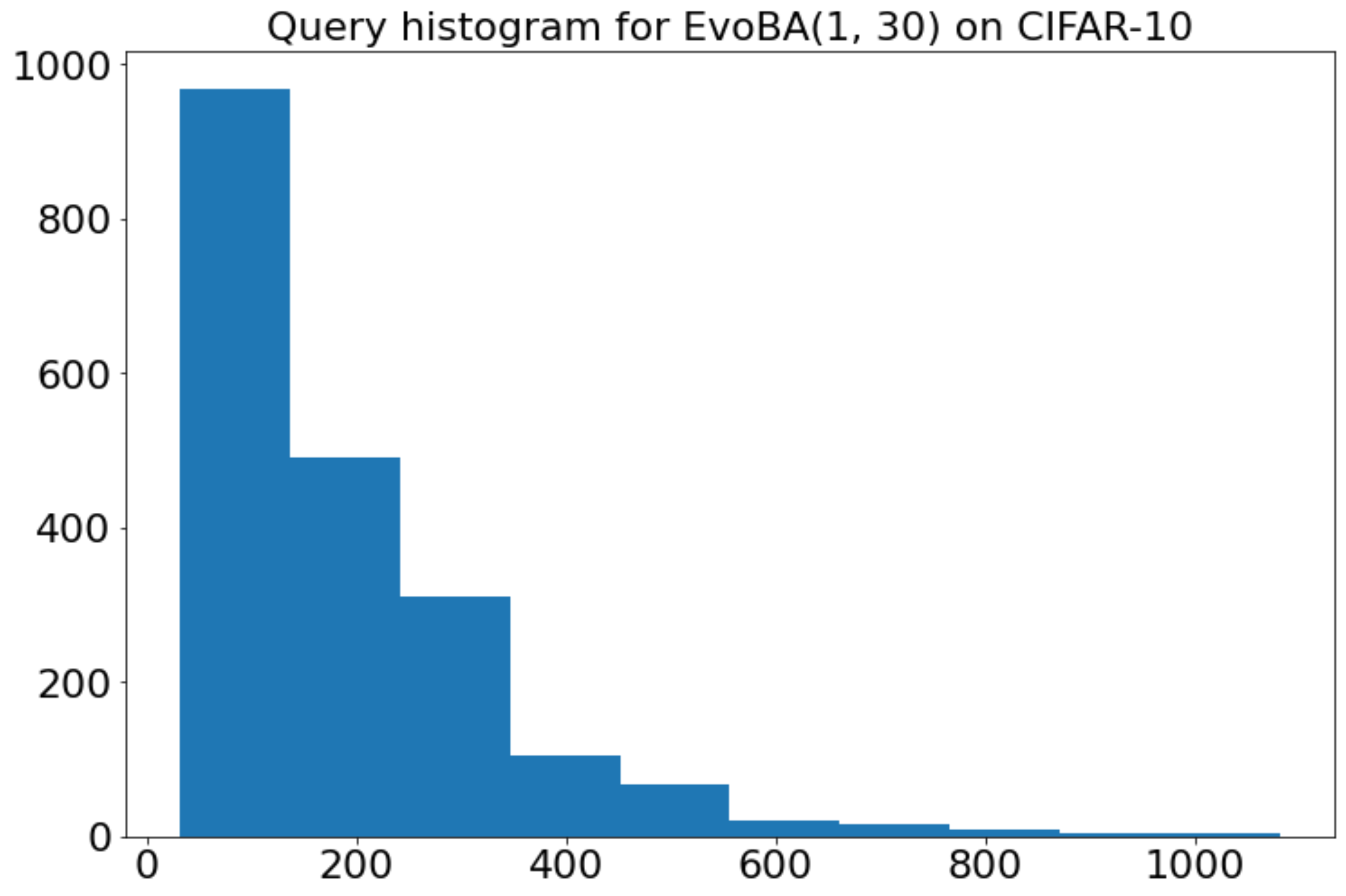}
    \caption{Histogram of query counts obtained by \textbf{EvoBA(1,30)} on CIFAR-10 with target model ResNet-50. The distribution is very heavy on small values, with few outliers.}
    \label{fig:histo_queries_evoba_cifar10}
\end{figure}

If we restrict $\textbf{EvoBA(1,10)}$ to its top $56\%$ perturbed images in terms of query-efficiency, it achieves an average of $192.79$ queries, which is below $\textbf{SimBA-LCB}$'s queries average of $196.86$. Similarly, if we restrict $\textbf{EvoBA(1,10)}$ to its top $56\%$ perturbed images in terms of $L_2$-efficiency, it achieves an $L_2$ of $3.07$, which is higher, but closer to $\textbf{SimBA-LCB}$'s $L_2$ result of $2.37$.

$\textbf{CompleteRandom}$ achieves a success rate of $60.5\%$, far below $\textbf{EvoBA}$'s $100\%$, but surprisingly above $\textbf{SimBA-LCB}$'s $56\%$. However, the nature of $\textbf{CompleteRandom}$'s perturbations is to lie at high distances, achieving $L_0$ and $L_2$ distances that are significantly higher than the distances achieved by the other methods.

\begin{figure}[ht]
    \centering
    \includegraphics[width=0.9\columnwidth]{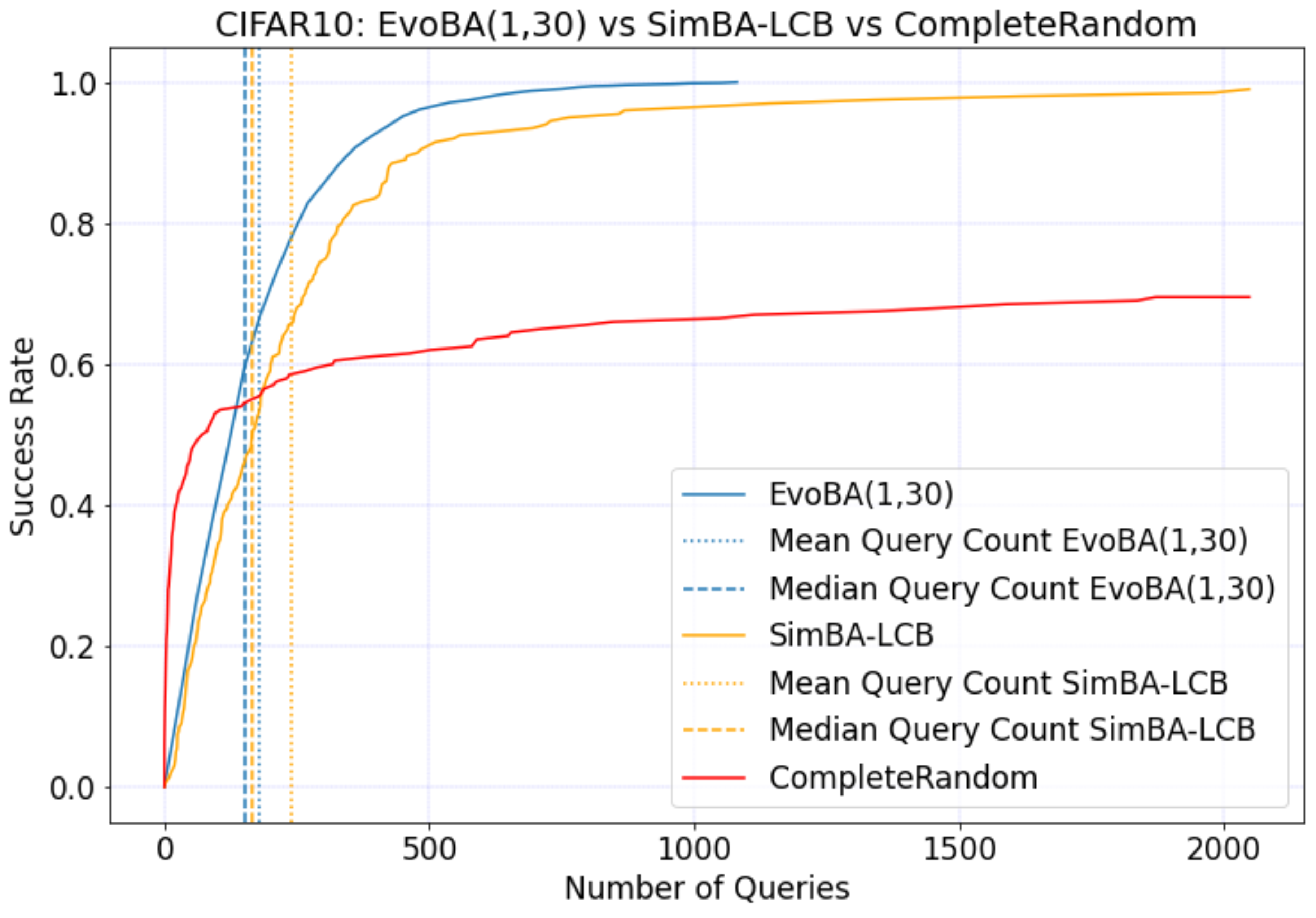}
    \caption{The success rate (ratio of perturbed images) as a function of the maximum query budget. We compare $\textbf{EvoBA}$ with the strong baseline $\textbf{SimBA}$ and with $\textbf{CompleteRandom}$ on CIFAR-10.}
    \label{fig:success_rate}
\end{figure}
\subsection{Results on CIFAR-10}
\begin{figure*}[!h]
    \centering
    \includegraphics[width=0.99\textwidth]{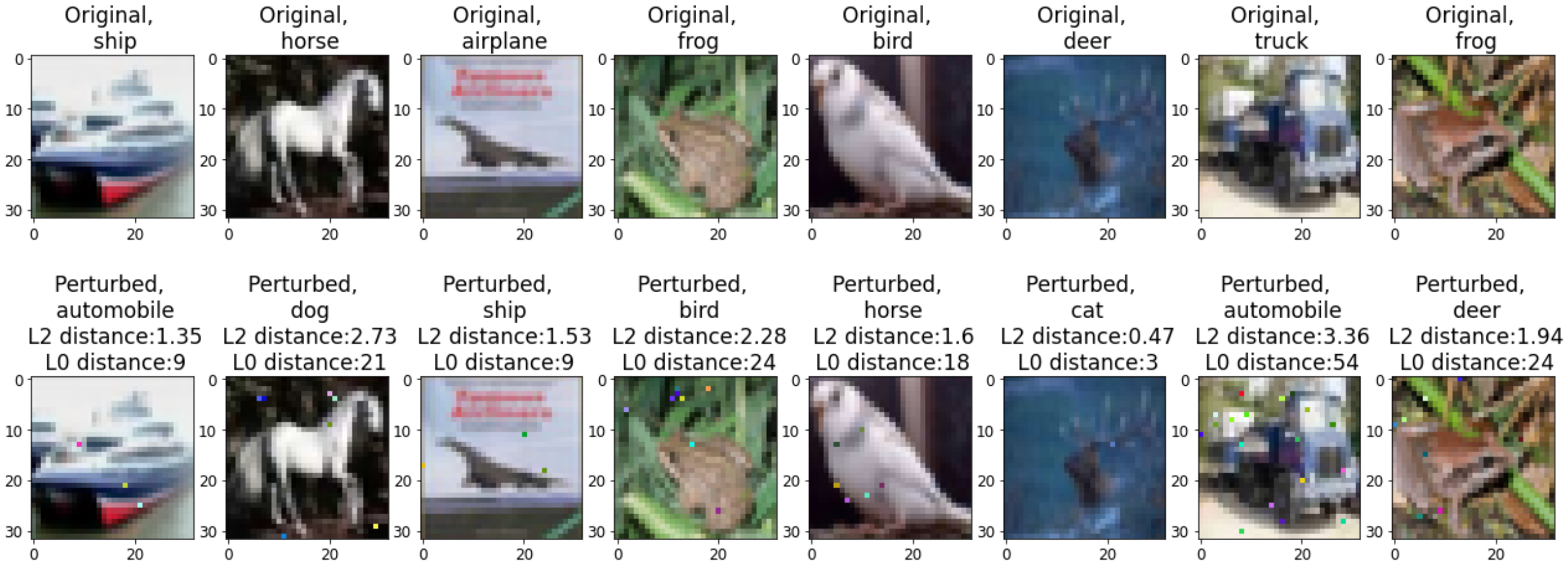}
    \caption{The first row contains original CIFAR-10 samples, which are classified correctly by ResNet-50. The second row contains adversarial examples created by $\textbf{EvoBA}$, and are labelled with the corresponding ResNet-50 predictions. Furthermore, we also provide the $L_2$ and $L_0$ distances between the unperturbed and perturbed samples. }
    \label{fig:band_cifar10}
\end{figure*}

We impose an $L_0$ perturbation limit of $100$, and a query threshold of $2000$ for both $\textbf{EvoBA}$ and $\textbf{CompleteRandom}$. We randomly sample $2000$ images for $\textbf{EvoBA}$ and $50$ images for $\textbf{SimBA-LCB}$. $\textbf{EvoBA}$ and $\textbf{SimBA-LCB}$ are run on the exact same target ResNet-50 model, while $\textbf{SimBA-DCT}$ and $\textbf{SimBA-CB}$ also run on a target ResNet-50 model. $\textbf{AutoZOOM-BiLIN}$ targets an InceptionV3 model.

$\textbf{SimBA-LCB}$ required an average of $\textbf{26.15s}$ per CIFAR-10 sample, while $\textbf{EvoBA}$ required $\textbf{1.91s}$ per sample.

\begin{table}[h]
    \centering
    \begin{tabular}{ c c c c c} 
 \hline
  & SR & QA & L0 & L2 \\ [0.5ex] 
 \hline
 \textbf{EvoBA(1,30)} & 100\% & 178.56 & 17.67 & 1.82 \\ 
 \textbf{SimBA-LCB} & 100\% & 206.5 & 99.46 & 1.73 \\
 \textbf{SimBA-CB} & 100\% & 322 & - & 2.04 \\
 \textbf{SimBA-DCT} & 100\% & 353 & - & 2.21 \\
 \textbf{AutoZOOM-BiLIN} & 100\% & 85.6 & - & 1.99 \\ 
 \textbf{CompleteRandom} & 69.5\% & 161.2 & 97.17 & 3.89 \\ [0.1ex] 
 \hline
    \end{tabular}
    \caption{CIFAR-10 results. \textbf{AutoZOOM}, \textbf{SimBA-CB}, and \textbf{SimBA-DCT} do not report the $L_0$ metrics. However, we have discussed already why it is very likely that \textbf{AutoZOOM} perturbs most of the pixels. }
    \label{tab:my_label2}
\end{table}

All the attacks achieved $100\%$ success rate in the CIFAR-10 experiments, with the sole exception of $\textbf{CompleteRandom}$, which only got $69.5\%$. $\textbf{EvoBA(1,30)}$ has a better query average when compared to all the $\textbf{SimBA}$ approaches, which targeted the same ResNet-50 architecture. While $\textbf{EvoBA(1,30)}$ targeted a different architecture than $\textbf{AutoZOOM-BiLIN}$, we still remark how the latter is twice more query efficient. However, $\textbf{EvoBA(1,30)}$ surprisingly achieves an $L_2$ metric which is better than the reported numbers of $\textbf{SimBA-CB}$ and $\textbf{SimBA-DCT}$, which are $L_2$ adversarial attacks.

\begin{figure}[ht]
    \centering
    \includegraphics[width=0.8\columnwidth]{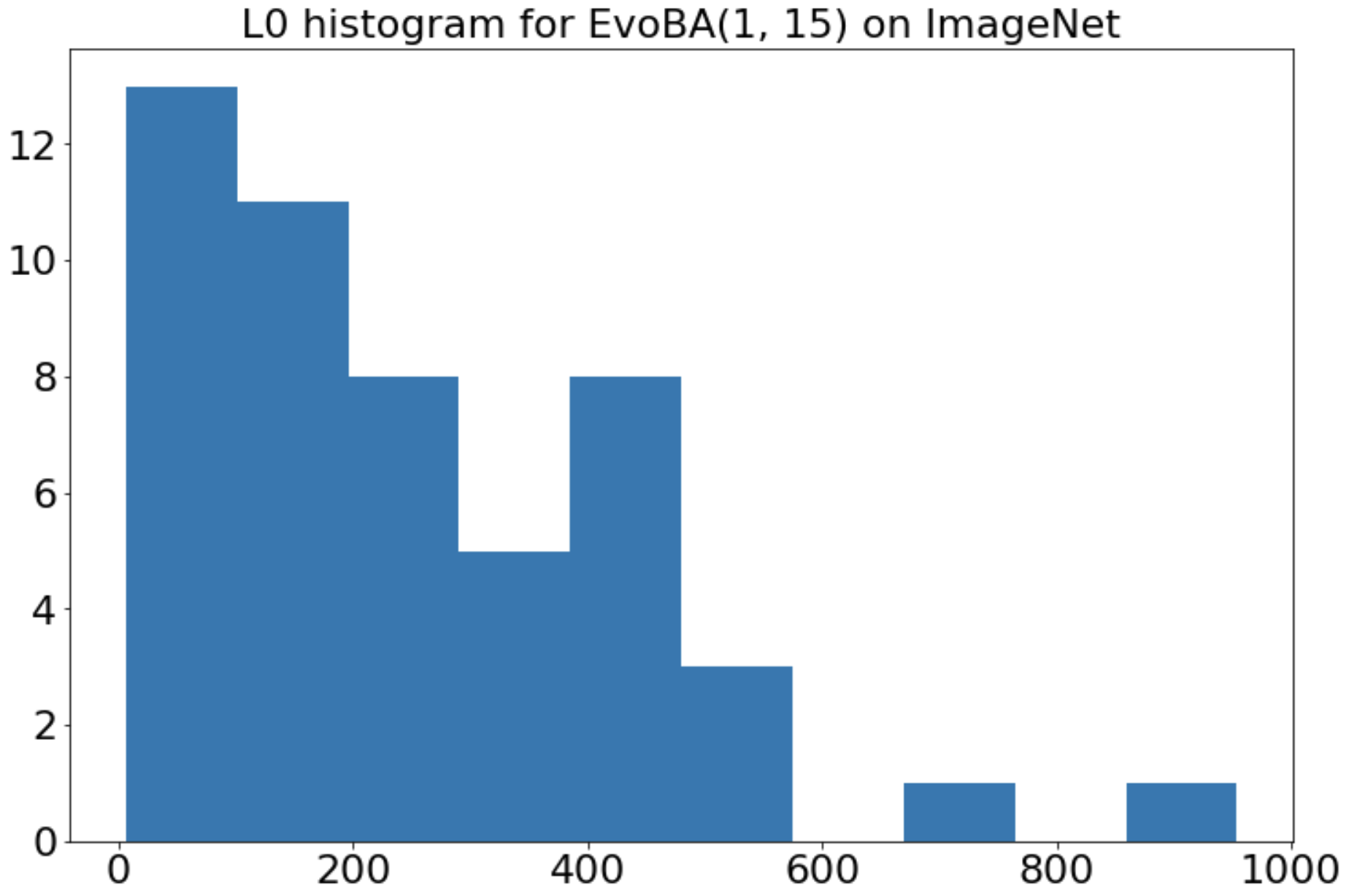}
    \caption{Histogram of $L_0$ perturbation norms obtained by $\textbf{EvoBA(1,15)}$ on ImageNet with target model ResNet-50. The distribution is very heavy on small values, with few outliers.}
    \label{fig:histo_l0_evoba}
\end{figure}

\begin{figure}[ht]
    \centering
    \includegraphics[width=0.8\columnwidth]{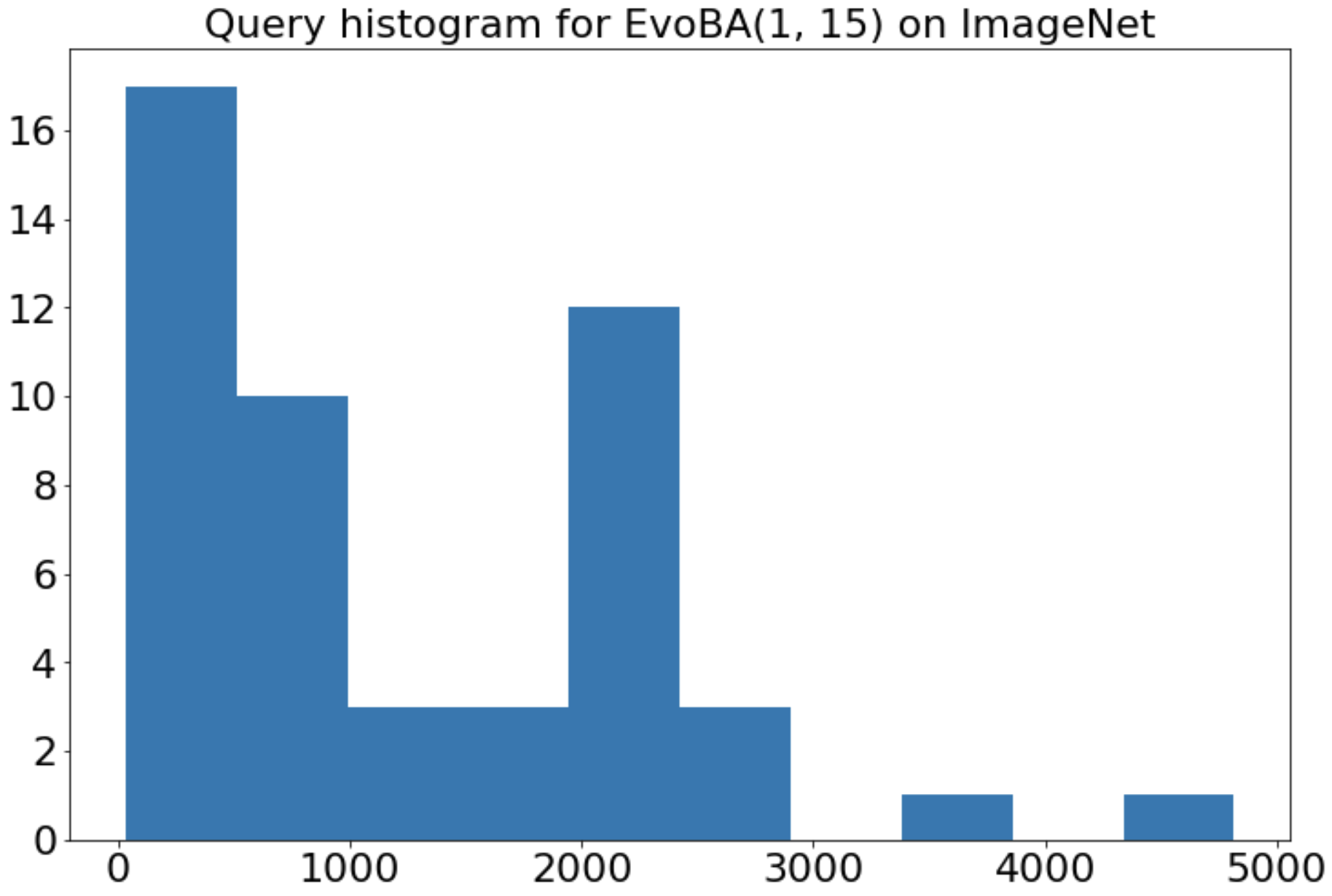}
    \caption{Histogram of query counts obtained by \textbf{EvoBA(1,15)} on ImageNet with target model ResNet-50. The distribution is very heavy on small values, with few outliers.}
    \label{fig:histo_query_evoba}
\end{figure}

It is not as much of a surprise the fact that $\textbf{EvoBA(1,30)}$ achieves a considerably better $L_0$ metric when compared to $\textbf{SimBA-LCB}$ ($17.67$ vs $99.46$).

In Figures \ref{fig:histo_l0_evoba_cifar10} and \ref{fig:histo_queries_evoba_cifar10} we plot the histograms of the $L_0$ perturbation norms, respectively of the query counts obtained by $\textbf{EvoBA(1,30)}$. Both are highly skewed towards low values, showing how $\textbf{EvoBA}$ does well in finding quick small perturbations with respect to the $L_0$ norm.

The success rate of $\textbf{CompleteRandom}$ ($69.5\%$) and its low average query count ($161.2$) are surprisingly good results for the trivial nature of the method, outlining once again the lack of robustness in complex image classifiers. However, these come at the cost of an average $L_0$ that is roughly $5.5$ times higher and of an average $L_2$ that is roughly $1.4$ times higher in comparison with $\textbf{EvoBA(1,30)}$'s average results.

In Figure \ref{fig:success_rate} we compare $\textbf{EvoBA(1,30)}$ with $\textbf{SimBA-LCB}$, while also providing the $\textbf{CompleteRandom}$ results. We plot the success rate as a function of the number of queries in order to understand how each method behaves for different query budgets. $\textbf{EvoBA(1,30)}$ has a better success rate than $\textbf{SimBA-LCB}$ for any query budget up to $2000$. For very low query budgets (under $112$ queries), $\textbf{CompleteRandom}$ has a better success rate than $\textbf{EvoBA(1,30)}$, but it starts converging fast after their intersection point to the success rate of $69.5\%$. It is natural for the $\textbf{CompleteRandom}$ strategy to find quick perturbations for the least robust images, as it performs bulk perturbations of many pixels at once, while $\textbf{EvoBA}$ does all perturbations sequentially. This illustrates the natural trade-off between exploration and exploitation that any black-box optimization problem encounters.

\subsection{Results on ImageNet}
We adopt a similar framework to $\textbf{AutoZOOM}$: we randomly sample 50 correctly classified images and run $\textbf{EvoBA}$ on top of them. For $\textbf{EvoBA}$, similarly to $\textbf{SimBA}$, we use a ResNet50 model, while $\textbf{AutoZOOM}$ uses an InceptionV3. We impose an $L_0$ perturbation limit of $1000$, and a query threshold of $10000$. The $L_0$ limit we impose is reasonable, as each time we perturb a pixel we actually modify all of its three channels, therefore the $1000$ limit stands for approximately $0.66\%$ of the image pixels.

\begin{table}[h]
    \centering
    \begin{tabular}{ c c c c c} 
 \hline
  & SR & QA & L0 & L2 \\ [0.5ex] 
 \hline
 \textbf{EvoBA(1,15)} & 100\% & 1242.4 & 247.3 & 6.09 \\ 
 \textbf{EvoBA(1,20)} & 100\% & 1412.51 & 211.03 & 5.72 \\
 \textbf{SimBA-CB} & 98.6\% & 1665 & - & 3.98 \\
 \textbf{SimBA-DCT} & 97.8\% & 1283 & - & 3.06 \\
 \textbf{AutoZOOM-BiLIN} & 100\% & 1695.27 & - & 6.06 \\ [0.1ex] 
 \hline
    \end{tabular}
    \caption{ImageNet results. \textbf{AutoZOOM}, \textbf{SimBA-CB}, and \textbf{SimBA-DCT} do not report the $L_0$ metrics. However, we have discussed already why it is very likely that \textbf{AutoZOOM} perturbs most of the pixels. }
    \label{tab:my_label3}
\end{table}

\begin{figure*}[ht]
    \centering
    \includegraphics[width=0.99\textwidth]{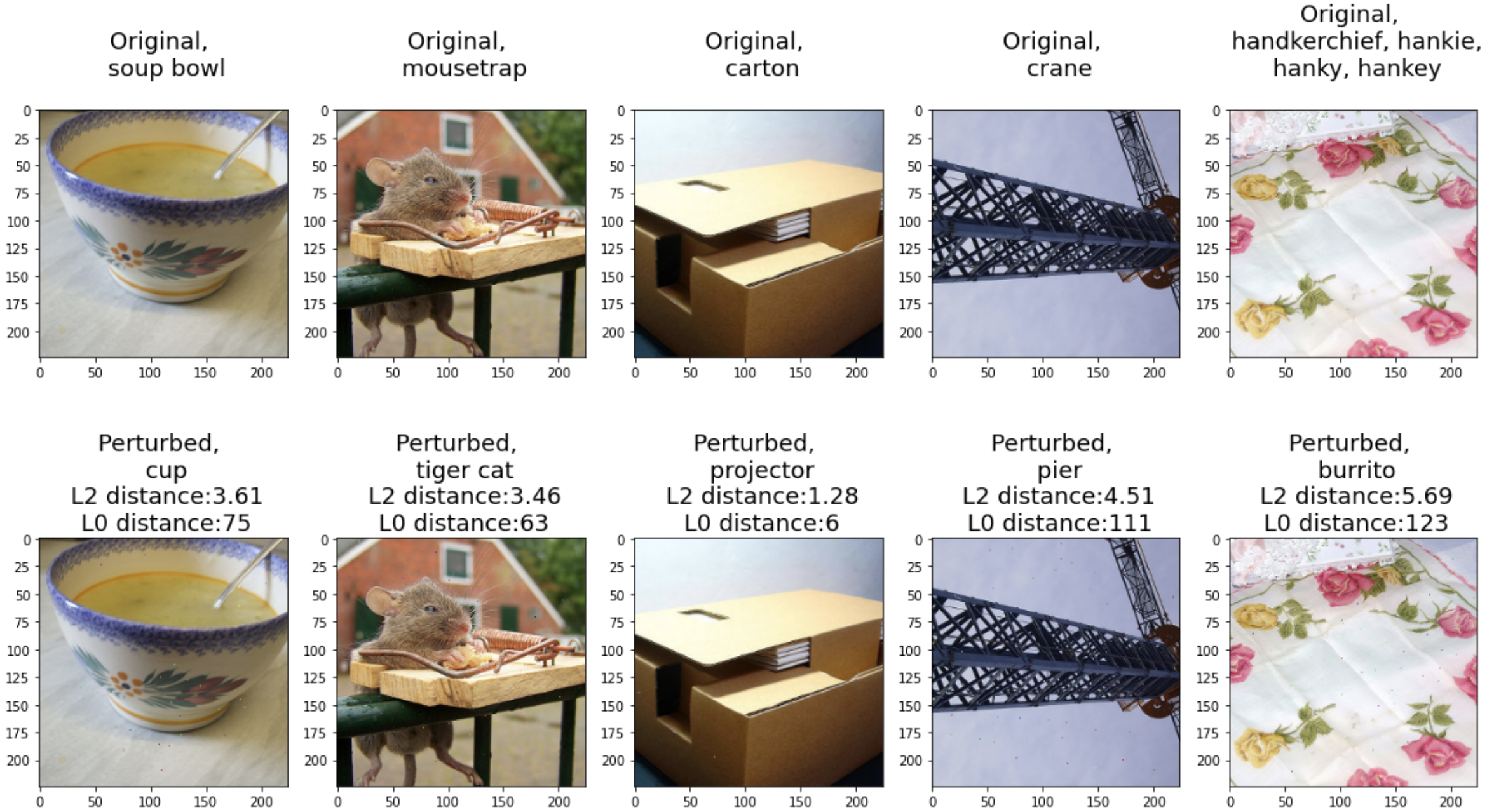}
    \caption{The first row contains original ImageNet samples, which are classified correctly by ResNet-50. The second row contains adversarial examples created by $\textbf{EvoBA}$, and are labelled with the corresponding ResNet-50 predictions. Furthermore, we also provide the $L_2$ and $L_0$ distances between the unperturbed and perturbed samples. }
    \label{fig:band}
\end{figure*}
The median number of queries for $\textbf{EvoBA(1,15)}$ is surprisingly low: $728.5$. Its median $L_0$ is $200$, and its median $L_2$ is $5.69$. This shows once more how we met our goal to minimise the query count and $L_0$ perturbation, as the medians are significantly smaller than the average values, showing how the distributions are biased towards low values. In comparison, the $L_2$ mean and median are close, indicating that the good $L_2$ results we get are a consequence of our other constraints rather than an actual objective.

$\textbf{EvoBA(1,15)}$ achieves the best average query metric among the given experiments. Surprisingly, its $L_2$ is almost equal to the one of $\textbf{AutoZOOM-BiLIN}$. $\textbf{EvoBA(1,15)}$ also achieves a $100\%$ success rate, which is in line with $\textbf{AutoZOOM-BiLIN}$, and better than $\textbf{SimBA}$'s results.

\subsection{Qualitative results}
In Figures \ref{fig:band_cifar10} and \ref{fig:band}, we provide samples of CIFAR-10, respectively of ImageNet perturbed images, together with their initial and perturbed labels. The perturbations are almost imperceptible to the human eye, and they look mostly like regular noise. Considering the highly-biased nature of the $L_0$ histogram in Figures \ref{fig:histo_l0_evoba_cifar10} and \ref{fig:histo_l0_evoba} towards small values, it is natural to expect well-crafted perturbations.


\section{Conclusion} \label{sect:concl}

We proposed $\textbf{EvoBA}$, an $L_0$ black-box adversarial attack based on an evolution strategy, which serves as a powerful baseline for black-box adversarial attacks.
It achieves results in line with state-of-the-art approaches, such as $\textbf{AutoZOOM}$, but is far less complex. Simple yet efficient methods, such as $\textbf{EvoBA}$, $\textbf{SimBA}$, $\textbf{DeepSearch}$, and even $\textbf{CompleteRandom}$ shed a light on the research potential of the black-box adversarial field, outlining the inherent security issues in many machine learning applications.

We plan to aggregate $\textbf{EvoBA}$ with the other strong black-box baseline attacks we have seen for different norms ($\textbf{SimBA}$ for $L_2$ and $\textbf{DeepSearch}$ for $L_{\infty}$) and create a generic open-source framework to empirically assess the robustness of image classifiers and to help with their development process. Such a toolbox could also be used to assess the quality of adversarial training methods, as many focus on improving the robustness of the target models with respect to $L_p$ norms.

Another immediate extension of our work is to customise $\textbf{EvoBA}$ to also support the targeted scenario, which can be done by adapting the fitness function used in the evolution strategy. This requires minimal changes in the implementation. $\textbf{EvoBA}$ can also be modified to make explicit use of its parallelism by distributing the generation evaluation steps to multiple workers.

$\textbf{EvoBA}$ could be easily extended for classification tasks in other fields, such as natural language processing. One would have to come with the right perturbation scheme in our approach, but we leave this for future research.

\newpage
\bibliographystyle{splncs04}
\bibliography{paper}

\appendix

\section{Consistency of the results}

As our attack is stochastic and relies heavily on a randomness generator, we did $10$ independent runs of $\textbf{EvoBA}$ on a sample of $200$ CIFAR-10 images.

The $\textbf{EvoBA}$ results from Table $\ref{tab:repeat_evoba}$ show how robust our method is. The query standard deviation is only $4.87$ (roughly $3\%$ of the query average $178.29$). Similarly, the standard deviation is very small for $\textbf{EvoBA}$ across all metrics.

The time required to perturb one CIFAR-10 sample was $1.91s$ on average, with a standard deviation of $0.1s$.

\begin{table}[h!]
    \centering
    \begin{tabular}{ c c c c c c c c c c c c c} 
 \hline
  \textbf{EvoBA}& Run 1 & Run 2 & Run 3 & Run 4 & Run 5 & Run 6 & Run 7 & Run 8 & Run 9 & Run 10 & \textbf{Mean} & \textbf{Std}\\ [0.5ex] 
 \hline
 \textbf{SR} & 100\% & 100\%& 100\%& 100\%& 100\%& 100\%& 100\%& 100\%& 100\%& 100\%& 100\%& 0 \\ 
 \textbf{L0} & 17.37 & 18.29& 17.94& 17.31& 17.27& 17.33& 17.76& 17.77& 18.43& 16.89& 17.63& 0.489 \\
 \textbf{L2} & 1.81 & 1.87 & 1.86 & 1.83 & 1.83 & 1.82 & 1.81 & 1.84 & 1.87 & 1.79 & 1.83 & 0.027\\
 \textbf{Q} & 175.6 & 184.9 & 181.3 & 174.7 & 174.55 & 175.45 & 179.8 & 184.5 & 186.1 & 170.95 & 178.285 & 4.871 \\ [0.1ex] 
 \hline
    \end{tabular}
    \caption{Ten independent random runs of $\textbf{EvoBA}$.}
    \label{tab:repeat_evoba}
\end{table}

\section{More baselines}

We show how $\textbf{EvoBA}$ compares with other well-known $L_2$ black-box adversarial attacks which work under the query-limited scenario. For this, we reuse the results from \cite{guo2019simple}, where the same model (ResNet-50) is targeted, and extend the metrics we presented in Table \ref{tab:my_label3}. The methods are $\textbf{QL-attack}$, introduced in \cite{ilyas2018black}, and $\textbf{Bandits-TD}$, introduced in \cite{ilyas2018prior}.

\begin{table}[ht!]
    \centering
    \begin{tabular}{ c c c c c} 
 \hline
  & SR & QA & L0 & L2 \\ [0.5ex] 
 \hline
 \textbf{EvoBA(1,15)} & 100\% & 1242.4 & 247.3 & 6.09 \\ 
 \textbf{EvoBA(1,20)} & 100\% & 1412.51 & 211.03 & 5.72 \\
 \textbf{SimBA-CB} & 98.6\% & 1665 & - & 3.98 \\
 \textbf{SimBA-DCT} & 97.8\% & 1283 & - & 3.06 \\
  \textbf{QL-attack} & 85.4\% & 28174 & - & 8.27 \\
   \textbf{Bandits-TD} & 80.5\% & 5251 & - & 5.00 \\
 \textbf{AutoZOOM-BiLIN} & 100\% & 1695.27 & - & 6.06 \\ [0.1ex] 
 \hline
    \end{tabular}
    \caption{ImageNet results which show how $\textbf{EvoBA}$ is comparable with $L_2$ attacks from the $L_2$ distance point of view, while being an $L_0$ attack.}
    \label{tab:simba_paper_results}
\end{table}







\end{document}